%% file: main.tex
\documentclass[9pt,twocolumn,twoside]{pnas-new}

\articletype{Social Sciences}

\templatetype{pnasresearcharticle} % Choose template
\usetikzlibrary{arrows.meta,positioning,calc}

\begin{document}

\title{Synthetic Participants for Real Interventions: The Limits of Augmenting Behavioral RCTs with Fine-Tuned LLMs}

\author[a]{Tommaso Batistoni}
\author[a]{Martina Beretta}
\author[a]{Raymond Low}
\author[b]{Ariele Niccoli}
\author[a,1]{Raymond Duch}

\affil[a]{University of Oxford} 
\affil[b]{University of Florence}

\leadauthor{Batistoni}

\significancestatement{Can large language models (LLMs) forecast the effect of a behavioral intervention before a randomized trial is run, or sharpen estimates from a small human sample? We test both uses on two field experiments that raised COVID-19 vaccination in Ghana and Sweden, with models used as released, fine-tuned on related trials and surveys, fine-tuned on a pilot of the trial, or both. Trained on the pilot, the models reproduce each arm's vaccination rate and, where the questionnaire records attitudes, who is likely to vaccinate. They do not reproduce how the interventions changed behavior, and improve precision no more than a four-item regression. Imitating observed behavior is not the same as predicting how people respond to an intervention.

}

\authorcontributions{T.B., M.B., A.N., and R.D. designed research; T.B., M.B., and R.L. performed research; T.B. and R.D. analyzed data; and T.B., M.B., A.N., and R.D. wrote the paper.}
\authordeclaration{The authors declare no competing interest.}
\correspondingauthor{\textsuperscript{1}To whom correspondence should be addressed. E-mail: raymond.duch@nuffield.ox.ac.uk}


\begin{abstract}
Large language models could help experimentalists anticipate how participants will respond to an intervention before a trial is run, and sharpen the trial's estimates afterward. We evaluate both uses on two field experiments that promoted COVID-19 vaccination, one in rural Ghana and one in Sweden. We simulate each trial's participants with GPT-4o and with four open-weight models, which we use as released and after fine-tuning on \textit{(i)} a corpus of related trials and surveys, \textit{(ii)} a pilot sample from the trial itself, or \textit{(iii)} both in sequence. Each combination of trial, model and training protocol is evaluated on the recovery of treatment effects, on outcome rates within arms, on the accuracy of individual predictions, and on the precision gained when predictions are combined with human data. Released and corpus-tuned models misestimate vaccination rates by up to 75 percentage points and produce treatment effects that vary erratically in sign and size. Fine-tuning on the pilot brings the rates within sampling error of the observed ones, but the synthetic treatment effects follow the pilot's own contrasts rather than those of the held-out sample, and their errors cannot be distinguished from those of forecasts that use no model at all. Only in Sweden, where the questionnaire records attitudes toward vaccination, do pilot-tuned models predict individual decisions and raise precision, by up to 15\%; a logistic regression on the same attitude items achieves comparable gains. Models trained on quantitative data can thus reproduce observed outcomes without learning how interventions change them.
\end{abstract}

\dates{This manuscript was compiled on \today}
\doi{}

\maketitle
\thispagestyle{firststyle}
\ifthenelse{\boolean{shortarticle}}{\ifthenelse{\boolean{singlecolumn}}{\abscontentformatted}{\abscontent}}{}

\dropcap{R}andomized controlled trials (RCTs) are the standard of causal inference in the social and health sciences, but their cost, duration and logistical demands limit the questions researchers can afford to ask. Large language models (LLMs) may relax these constraints. A small human sample paired with a larger set of model predictions could yield treatment-effect estimates as precise as those of a much larger trial, at a fraction of the cost and time \citep{Angelopoulosetal2023, Angelopoulosetal2024, Broskaetal2025}. Even before any human data are collected, synthetic participants could forecast the outcome in each arm, anticipate treatment effects and indicate who is most likely to respond. Researchers could use such forecasts to refine instruments and interventions, to assess statistical power under alternative designs and to screen hypotheses before committing resources to the field \citep{Ludwigetal2026}. All of these uses require that synthetic predictions align with the behavior of the target population in the setting of interest.
The evidence on this alignment is mixed. Frontier models reproduce aggregate behavioral patterns in some domains \citep{Hewittetal2024, Parketal2024, Cuietal2025} and fail systematically in others. Their predictions over-represent US-centric demographic groups and perform poorly for populations that are under-represented in pretraining corpora \citep{Santurkaretal2023, Caoetal2025}. Synthetic responses vary less than human ones and yield different regression coefficients on the same covariates \citep{Bisbeeetal2024}. Individual predictions do not outperform demographic baselines and overstate how far demographics determine attitudes \citep{Chenetal2026, Stojanov2026}. Even where models recover the main experimental effects, they may exaggerate their size \citep{Cuietal2025, Hewittetal2024}. Most of this evidence comes from online experiments with text-based treatments and attitudinal outcomes, or from correlations between predicted and observed effects pooled across many studies. An experimentalist faces a narrower question, namely whether synthetic predictions can be trusted for a particular intervention, population and behavioral outcome. Accuracy averaged across studies offers little guidance on that point.

We take the experimentalist's perspective. Models fine-tuned on large collections of experiments now predict, with considerable accuracy, the choices of new participants in familiar tasks and the distribution of answers under new conditions of familiar studies \citep{Binzetal2025, Kollurietal2025, Xieetal2025}. Much of that accuracy, however, is matched by models of about one billion parameters trained on the same data \citep{OhGobet2026}, and it is earned within families of closely related studies rather than on the effect of a new intervention in a new population. Rather than build such a general-purpose model, we ask whether LLMs can be specialized to a particular trial with data an experimentalist could plausibly obtain: a curated corpus of related trials and surveys, selected for its relevance to the behavioral domain, the population and the interventions, and a small pilot sample from the trial itself. This approach targets the alignment a given trial requires rather than alignment in general, needs modest data rather than a behavioral corpus of pretraining scale and, with parameter-efficient fine-tuning, is cheap enough to repeat for every new trial. Four training protocols isolate the contribution of each data source and mirror the situations an experimentalist might face: the released model (T0), fine-tuning on the corpus alone (T1), fine-tuning on the pilot alone (T2), and fine-tuning on the corpus and then on the pilot (T3). Each source has a long-standing counterpart in trial methodology. The corpus plays the part of historical controls, which strengthen a new trial only to the extent that the earlier studies are exchangeable with it \citep{Pocock1976, Vieleetal2014}. The pilot is the small internal sample whose effect estimates have long been known to be too noisy to guide even a power calculation \citep{Kraemeretal2006, Leonetal2011}. And augmenting a trial with predictions is the language-model counterpart of adjusting its estimate for a prognostic score trained on external data, a practice that regulators now accept \citep{Schuleretal2022, EMA2022PROCOVA}.

We evaluate each training protocol along four complementary dimensions: the recovery of the treatment effects observed in the human trial (D1); the agreement between synthetic and human outcome distributions within each treatment arm (D2); the accuracy of individual predictions (D3); and the gain in precision from augmenting human outcomes with synthetic predictions (D4). For D4 we use Predict-Then-Debias (PTD) inference, a form of prediction-powered inference that accommodates cluster randomization \citep{Angelopoulosetal2023, Klugeretal2025}. We report the dimensions separately, because success on one need not imply success on another and because their relative importance depends on whether the predictions are meant to inform the design of a trial or to augment its statistical inference.

We compare synthetic predictions with two completed field experiments on vaccination uptake in which vaccination was verified in administrative records: a cluster-randomized trial in rural Ghana \citep{Duchetal2023} and an individually randomized trial in Sweden \citep{CamposMercadeetal2021}. The two trials differ in population, in the unit of randomization and in their interventions, which range from informational nudges to cash incentives. Models, training corpus, prompts, hyperparameters and evaluation metrics were preregistered \citep{duch_batistoni_beretta_low_2026}, and departures from the preregistration are listed in SI Appendix, section 2.

Three findings emerge. First, without pilot data, synthetic participants are unreliable on most dimensions. Released and corpus-tuned models misestimate arm-level vaccination rates by up to 75 percentage points (pp), and their treatment effects vary widely in sign and size across model families, model sizes and target trials. Second, fine-tuning on the pilot makes the synthetic samples resemble the human ones, but the resemblance is borrowed from the pilot. Synthetic vaccination rates inherit the pilot's sampling error, and synthetic treatment effects follow the pilot's own contrasts: closely in Sweden, and in Ghana as a heavily damped copy that the models reproduce in the pilot's villages and barely carry to any other. Across all eight treatment contrasts, the errors of the pilot-tuned models lie within the range that sampling noise in the human benchmark produces by itself, and so do the errors of two forecasts that need no model: the pilot's own estimates and a forecast of no effect. Third, synthetic predictions improve statistical precision only where the participant's profile carries predictive information beyond the covariates the trial's analysis already uses. In Sweden, four questionnaire items on vaccine attitudes yield gains equivalent to adding up to 15\% more labeled participants, but a logistic regression on the same four items comes within 3 pp of the best model's gain in every arm. Reproducing observed behavior, in short, does not imply recovering causal effects, and closer predictive alignment need not add information to experimental inference.

\section*{Results}

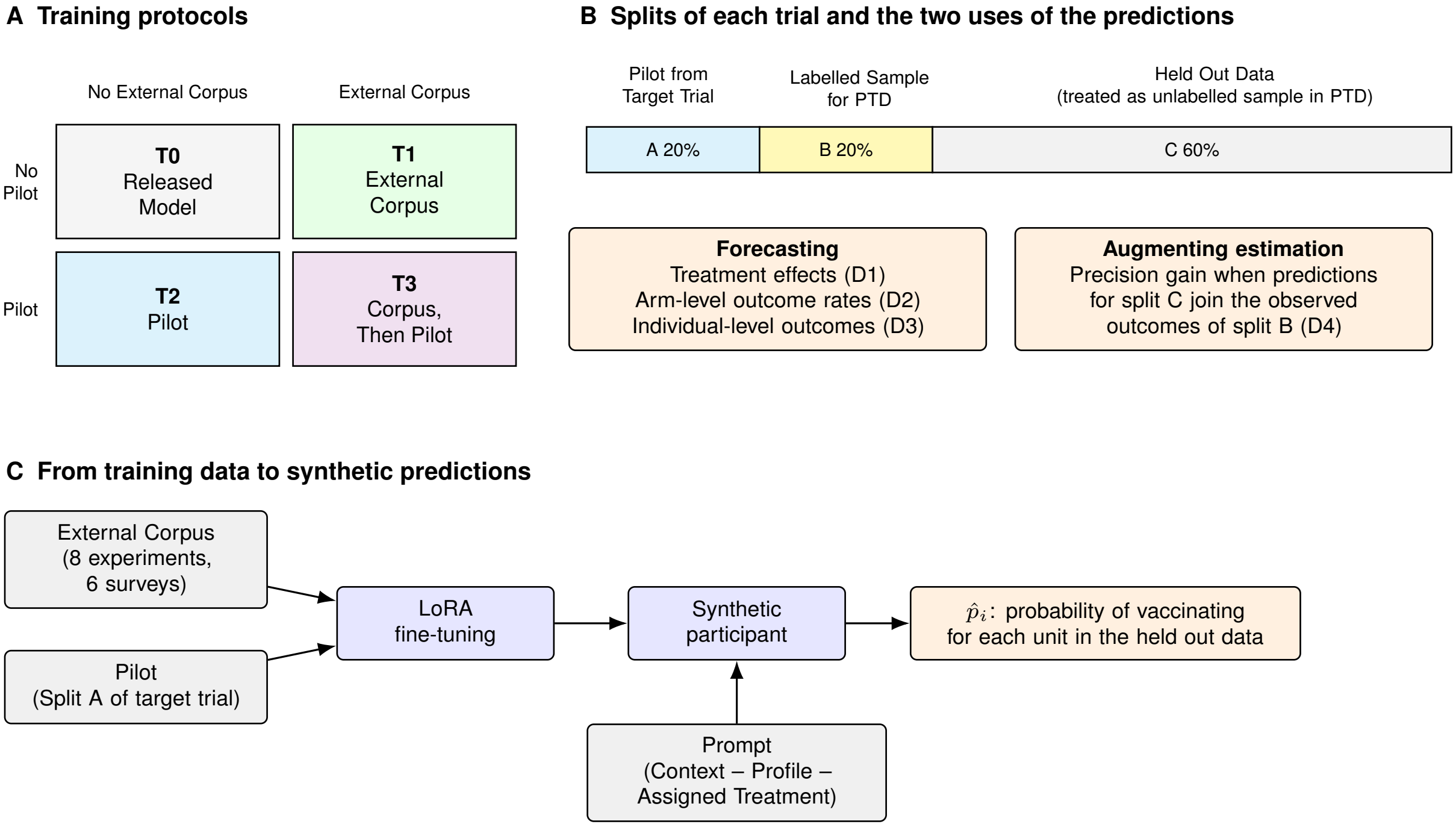
\begin{figure*}[t]
\centering
\input{Figures/Main/fig1_design}
\caption{\textbf{Study design.} (\textit{A}) Four open-weight models enter every training protocol (T0--T3) and GPT-4o enters T0 only, across two trials (34 combinations). The external corpus, used for training in T1 and T3, contains eight vaccination experiments and six surveys (87,680 records; SI Appendix, section 3). (\textit{B}) Each trial is partitioned, stratified by arm and at the village level in Ghana, into a pilot sample (A), used only for training; a labeled sample (B); and a held-out sample (C). Predictions are evaluated against held-out outcomes on treatment effects (D1), rates within arms (D2) and individual outcomes (D3), and are combined with labeled outcomes under PTD inference (D4). (\textit{C}) Each model reads one prompt per participant, which concatenates the study context, the participant's profile and the assigned treatment, and returns a probability of vaccination.}
\label{fig:design}
\end{figure*}

Fig.~\ref{fig:design} outlines the study design. Four open-weight models from two families and at two scales (OLMo~2 7B and 32B; Qwen3 8B and 32B) enter four training protocols (T0--T3), and GPT-4o enters as released (T0 only), giving 34 combinations of trial, model and training protocol. Each trial is divided into a pilot sample (20\% of units), a labeled sample (20\%) and a held-out sample (60\%). The model reads one standardized prompt per participant, describing the trial's setting, the participant and the assigned treatment, and returns a probability of vaccination. The held-out outcomes are the primary benchmark against which the predictions are scored; under PTD inference they are also treated as unlabeled and combined with the outcomes of the labeled sample (SI Appendix, section 1). Analyses that were not preregistered are identified as exploratory (SI Appendix, section 9).

Re-estimating each trial's original specification on its held-out sample yields eight treatment contrasts, five in Sweden and three in Ghana. These contrasts are the benchmark, and we give each of them the same weight. In Sweden the SEK~200 payment raises first-dose vaccination by 4.7 pp (OR 1.32, 95\% CI 1.05--1.64), and the four other treatments shift it by between $-2.1$ and 0.8 pp. In Ghana the low and high cash incentives raise verified vaccination by 13.3 pp (OR 2.06, 95\% CI 1.02--4.19) and 11.2 pp, and the health message lowers it by 3.2 pp. Only the payment and the low cash incentive are statistically distinguishable from zero (SI Appendix, Table S1). The question is whether models can recover these contrasts and estimate them more precisely.

\subsection*{Without pilot data, models misestimate treatment effects and vaccination rates}

Released (T0) and corpus-tuned (T1) models perform poorly on all three forecasting dimensions. Their treatment effects differ sharply from one model to the next and at times reverse in sign: the four corpus-tuned models, for example, put the effect of the Swedish payment at 45, 20, 8 and $-5$ pp, against 4.7 pp in the held-out sample (Table~\ref{tab:d1}). Taken over all five Swedish contrasts, the root-mean-square error of the corpus-tuned effects is 8.0--19.1 pp, several times the 3.5 pp that sampling noise in the human benchmark can produce on its own (exploratory; Table~\ref{tab:summary}). In Sweden only one released open-weight model, OLMo~2 7B, comes close, and only because it predicts a vaccination rate of about 2\% in every arm and hence no effect at all. Released open-weight models misestimate arm-level vaccination rates by as much as 75 pp in Sweden and 57 pp in Ghana, and corpus tuning removes the largest of these errors without eliminating the rest (Fig.~\ref{fig:d2}). Individual predictions are no more informative. Neither training protocol predicts individual outcomes more accurately than the human vaccination rate of the participant's arm, with the exception of GPT-4o in Sweden, whose advantage is negligible (Brier skill score 0.02; Fig.~\ref{fig:d3}). In exploratory analyses, three of the four released open-weight models also predict that vaccination rises steeply with income, education, employment, altruism and prior COVID-19 infection. These associations are largely absent from the human data and suggest a stereotyped reading of the participant's profile (SI Appendix, section 9).

Two apparent successes complicate the picture. GPT-4o puts the Swedish payment effect at 4.2 pp, which matches the published estimate rather than the 4.7 pp observed in the held-out sample. Qwen3 32B estimates the Ghanaian low-cash effect to within 3 pp of the held-out benchmark under T0 and T1, is the only released model to reproduce the published ordering of the Ghanaian arms, and attains the smallest errors of the 17 Ghanaian combinations across the three contrasts (1.1 pp under T1 and 2.0 pp under T0; exploratory). Both models were trained after the trials were published, so exposure to their results is possible, but their training data cannot be inspected. Exposure does not guarantee reproduction, however: the Swedish trial appears in the released training data of OLMo~2, yet OLMo~2 does not reproduce its results (SI Appendix, section 8).

\subsection*{After pilot tuning, models follow the pilot rather than the trial}

Fine-tuning on the pilot (T2 and T3) makes the synthetic effects far less variable (Table~\ref{tab:d1}), and every one of them lies inside its preregistered 95\% prediction interval. This agreement is less informative than it appears, because the human benchmark is itself noisy. Refitting the human regression on bootstrap resamples of the held-out sample shows that sampling variation alone can produce a root-mean-square error of up to 3.5 pp across the Swedish contrasts and of up to 14.3 pp across the Ghanaian ones (97.5th percentiles; exploratory). The pilot-tuned models stay within these bounds, at 2.3--3.0 pp in Sweden and 6.1--13.3 pp in Ghana under T2. So, however, do two forecasts that require no model: the pilot's own estimates (1.7 and 11.9 pp) and a forecast of no effect in any arm (2.3 and 10.2 pp; Table~\ref{tab:d1}). At the scale of these trials, agreement with the benchmark cannot separate a model that has learned the effects from one that has learned nothing about them.

The pilot's own contrasts reveal what the models follow instead. In Sweden the pilot-only effects (T2) are no closer to the held-out effects than the pilot's own estimates, which they improve on in only 30\% of model--arm comparisons, and no closer than those of a logistic regression fitted to the pilot alone (exploratory; SI Appendix, section 9). The payment arm shows the pattern in miniature. Its raw contrast in the pilot is 1.6 pp, the pilot-only models estimate 1.6--2.8 pp, and the held-out effect is 4.7 pp.

In Ghana the pilot is a poorer guide, because with ten or eleven villages per arm its contrasts are dominated by chance differences between villages. It reverses the sign of the high-cash effect ($-6.9$ pp against 11.2 pp) and overstates the low-cash effect (22.4 pp against 13.3 pp). The pilot-only models learn these contrasts, but they store them in the identities of the pilot villages, which the prompt names, rather than in the treatment text, and carry little of them to other villages: among held-out participants the low-cash effect shrinks to 1.8--6.6 pp (Table~\ref{tab:d1}; SI Appendix, section 9). The shrinkage, perversely, improves the forecasts. Accuracy in Table~\ref{tab:d1} is summarized by the root-mean-squared error of a forecast across the three contrasts, and in Ghana the pilot's contrasts are so noisy that forecasting no effect at all beats them (RMSE 10.2 against 11.9 pp); models that shrink the pilot's contrasts toward zero therefore move toward the better of the two model-free forecasts. The corpus-then-pilot models, whose effects are the smallest of any training protocol, even edge past the zero forecast (RMSE 3.8--8.3 against 10.2 pp), but the improvement lies well within the range that sampling noise produces.

\begin{figure*}[t]
\centering
\includegraphics[width=\textwidth]{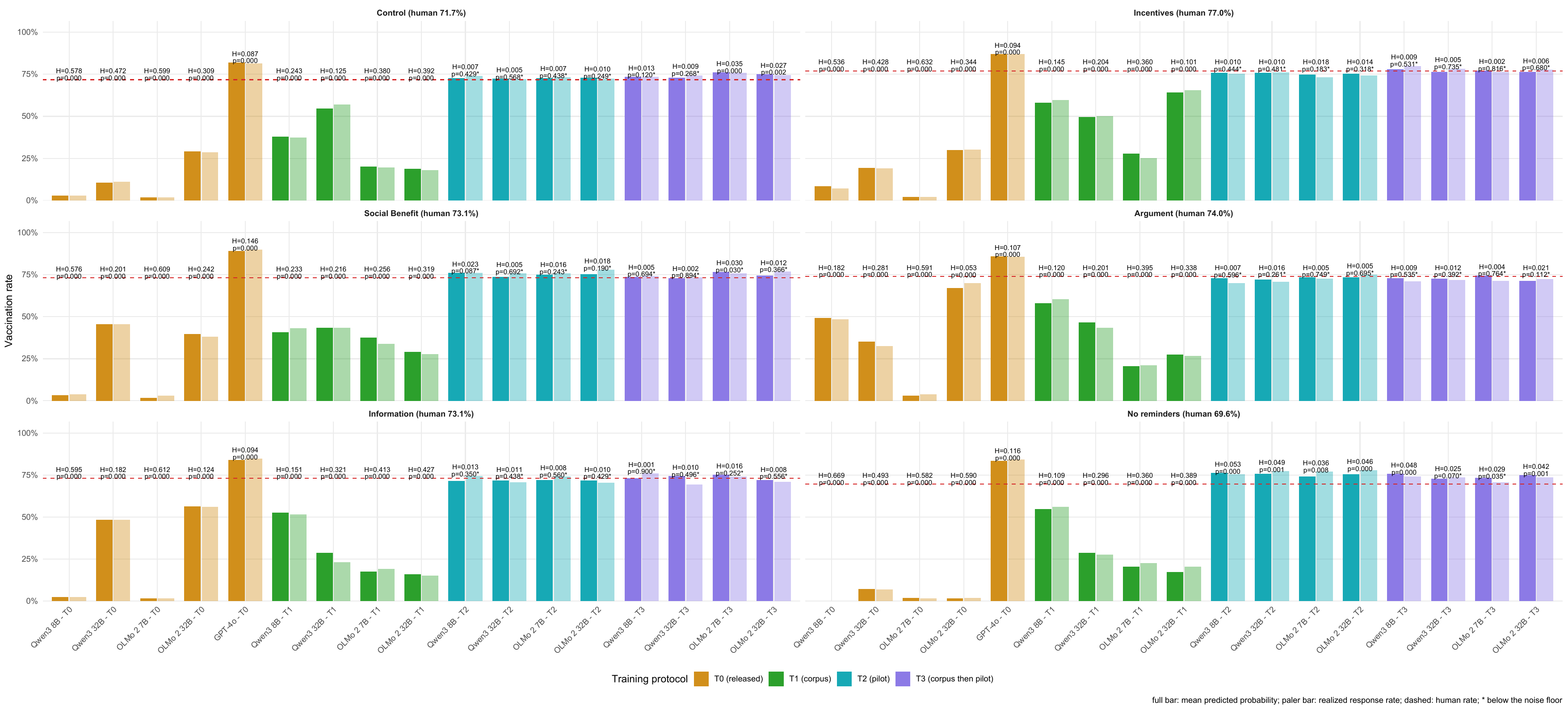}\\[2pt]
\includegraphics[width=0.75\textwidth]{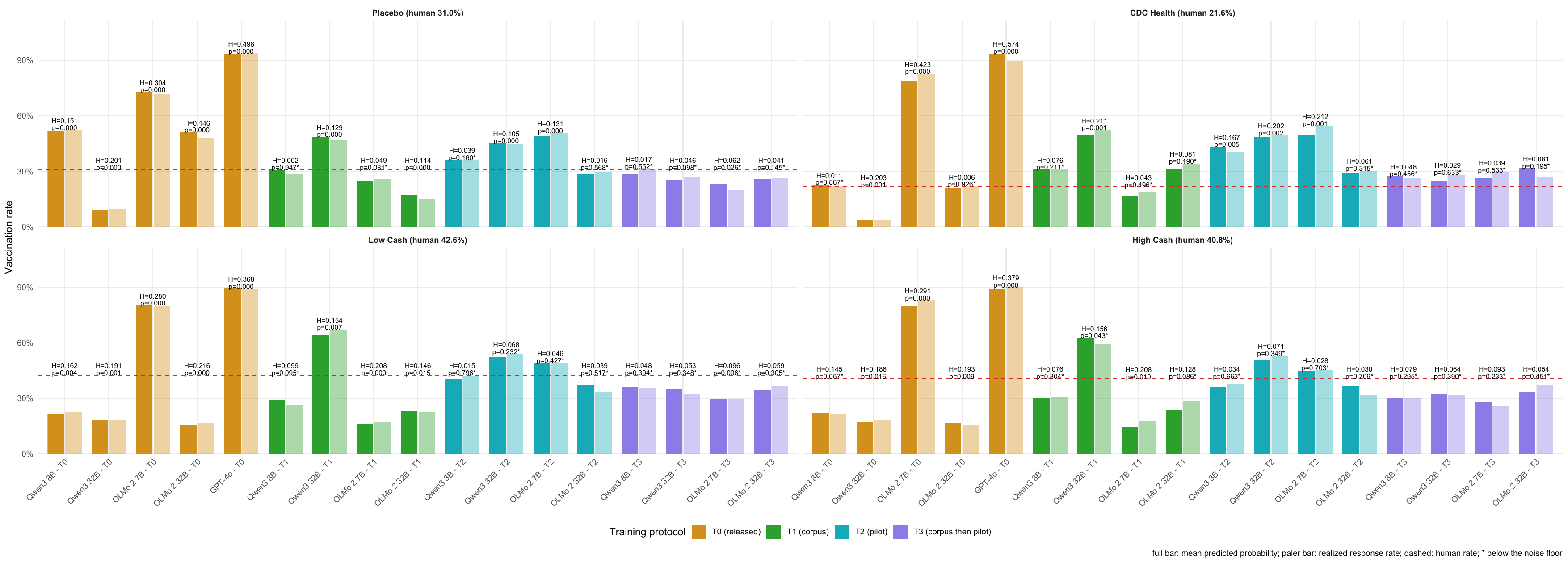}
\caption{\textbf{Arm-level vaccination rates (D2).} Mean predicted probability of vaccination (dark bars) against the held-out human rate (dashed line), by training protocol (rows) and arm (panels), for (\textit{A}) Sweden and (\textit{B}) Ghana. Annotations give the Hellinger distance between the synthetic and human Bernoulli distributions and the preregistered noise floor, the 97.5th percentile of the distance under bootstrap resampling of the human outcomes (participants in Sweden, villages in Ghana); asterisks mark distances below the floor. Light bars show the rate of realized Yes/No responses, for description only.}
\label{fig:d2}
\end{figure*}

The arm-level rates behind these contrasts appear far more accurate. We measure the distance between each synthetic and human rate by the Hellinger distance, a standard measure of the difference between two probability distributions, and compare it with a preregistered noise floor: the distance that resampling the human outcomes alone exceeds in only 2.5\% of resamples. More than 80\% of the pilot-tuned rates in each trial fall below this floor (Fig.~\ref{fig:d2}). But the accuracy, like the effects, is inherited from the pilot rather than learned. The pilot's own rates, used directly as a forecast, pass the same test in 83\% of the Swedish and 75\% of the Ghanaian arms (exploratory). And a logistic regression fitted to the same pilot reproduces about four-fifths of the variation in the Swedish pilot-only predictions; in neither trial are the pilot-tuned rates closer to the held-out rates than the regression's (exploratory; SI Appendix, section 9).

Training on the corpus and then on the pilot (T3) changes little in Sweden and somewhat more in Ghana, where the cluster-randomized pilot is noisiest. In Ghana agreement with the sign of the human effect rises from 7 to 10 of 12 model--arm comparisons, although the intervals of the two training protocols overlap in every arm; in Sweden it rises from 8 to 9 of 20. Only one of the 34 combinations, Qwen3 32B under T3 in Ghana, recovers both the sign and the statistical significance of every contrast, and even this combination understates the low-cash effect (10.7 pp). Because the released version of the same model already reproduces the published ordering of the Ghanaian arms, prior exposure cannot be ruled out.

Pilot tuning, then, makes the synthetic sample resemble the trial only to the extent that the pilot already does. The models reproduce the pilot's arm rates and contrasts, sampling error included, and add little information of their own.

\begin{table*}[t]
\centering
\caption{\textbf{Treatment effects (D1).} Average marginal effect of each arm against the control (pp) on the held-out sample: 4,972 participants in Sweden, and 1,350 participants with complete covariates in 137 villages in Ghana. Synthetic effects are the mean over 2,000 outcomes drawn from the predicted probabilities, each fitted with the original study's logistic specification; the human row fits the same regression to the observed outcomes ($^*$: $P < 0.05$, design-based standard errors), and the pilot row fits it to the pilot sample alone. RMSE is the root-mean-square difference from the human effects over the arms of the trial, every arm weighted equally (exploratory); sampling noise in the human benchmark produces an RMSE of up to 3.5 pp in Sweden and 14.3 pp in Ghana on its own. $^\dagger$: synthetic effect outside the preregistered joint 95\% prediction interval for a replication. Intervals and odds ratios for every entry are in SI Appendix, Table S1 and Fig.~S1; for Qwen3 8B under T0 in Sweden, where more than 1\% of fits separate, Table S1 also reports Firth-penalized estimates.}
\label{tab:d1}
\small
\begin{tabular}{llrrrrrrrrrr}
\toprule
 & & \multicolumn{6}{c}{Sweden} & \multicolumn{4}{c}{Ghana} \\
\cmidrule(lr){3-8}\cmidrule(lr){9-12}
 & & Payment & Social & Argu- & Infor- & No re- & RMSE & Health & Low & High & RMSE \\
 & & & impact & ments & mation & minders & & message & cash & cash & \\
\midrule
Human & held-out & 4.7$^*$ & 0.1 & 0.5 & 0.8 & $-$2.1 & -- & $-$3.2 & 13.3$^*$ & 11.2 & -- \\
Pilot estimate & no model & 2.6 & 1.0 & 0.2 & $-$1.7 & $-$0.5 & 1.7 & 0.8 & 22.4 & $-$6.9 & 11.9 \\
No effect & no model & 0.0 & 0.0 & 0.0 & 0.0 & 0.0 & 2.3 & 0.0 & 0.0 & 0.0 & 10.2 \\
\midrule
T0 & OLMo~2 7B & 0.2 & 0.0 & 1.1 & $-$0.1 & 0.0 & 2.3 & 3.9 & 4.2 & 4.7 & 7.6 \\
 & OLMo~2 32B & 0.5 & 9.4$^\dagger$ & 36.3$^\dagger$ & 27.3$^\dagger$ & $-$28.1$^\dagger$ & 23.5 & $-$34.1$^\dagger$ & $-$39.2$^\dagger$ & $-$36.1$^\dagger$ & 44.5 \\
 & Qwen3 8B & 5.7$^\dagger$ & 0.4 & 43.1$^\dagger$ & $-$0.4 & $-$2.9$^\dagger$ & 19.1 & $-$31.7$^\dagger$ & $-$34.9$^\dagger$ & $-$31.1$^\dagger$ & 40.6 \\
 & Qwen3 32B & 8.4$^\dagger$ & 33.8$^\dagger$ & 23.6$^\dagger$ & 37.8$^\dagger$ & $-$3.5 & 24.7 & $-$5.3 & 11.2 & 9.5 & 2.0 \\
 & GPT-4o & 4.2 & 5.7$^\dagger$ & 2.4 & 1.7 & 0.3 & 2.9 & $-$0.3 & $-$5.5$^\dagger$ & $-$3.9 & 14.0 \\
\addlinespace
T1 & OLMo~2 7B & 8.1 & 17.4$^\dagger$ & 0.4 & $-$1.7 & $-$0.5 & 8.0 & $-$13.1 & $-$13.0$^\dagger$ & $-$14.8$^\dagger$ & 22.1 \\
 & OLMo~2 32B & 45.1$^\dagger$ & 10.3$^\dagger$ & 8.9$^\dagger$ & $-$2.4 & $-$1.8 & 19.1 & 13.3 & 3.7 & 5.2 & 11.5 \\
 & Qwen3 8B & 20.0$^\dagger$ & 3.0 & 20.1$^\dagger$ & 14.7$^\dagger$ & 16.8$^\dagger$ & 15.3 & $-$1.2 & $-$2.9 & $-$2.2 & 12.2 \\
 & Qwen3 32B & $-$5.0$^\dagger$ & $-$11.2$^\dagger$ & $-$8.2$^\dagger$ & $-$25.6$^\dagger$ & $-$25.7$^\dagger$ & 17.6 & $-$1.3 & 13.5 & 11.3 & 1.1 \\
\addlinespace
T2 & OLMo~2 7B & 1.6 & 1.6 & 0.0 & $-$0.9 & 1.3 & 2.3 & 10.1 & 1.8 & $-$3.8 & 13.3 \\
 & OLMo~2 32B & 1.6 & 1.2 & $-$0.6 & $-$1.7 & 1.9 & 2.6 & $-$1.6 & 6.4 & 3.4 & 6.1 \\
 & Qwen3 8B & 2.5 & 2.5 & $-$0.4 & $-$1.6 & 3.2 & 3.0 & 5.7 & 4.8 & $-$5.4 & 11.9 \\
 & Qwen3 32B & 2.8 & 0.3 & $-$1.2 & $-$0.9 & 3.0 & 2.6 & 7.0 & 6.6 & 3.2 & 8.4 \\
\addlinespace
T3 & OLMo~2 7B & 0.9 & 0.3 & $-$1.7 & $-$1.5 & $-$2.2 & 2.2 & $-$0.7 & 5.2 & 4.2 & 6.3 \\
 & OLMo~2 32B & 0.9 & $-$0.8 & $-$3.9 & $-$3.2 & $-$0.4 & 3.3 & 0.1 & 3.5 & 3.3 & 7.5 \\
 & Qwen3 8B & 4.1 & $-$0.5 & $-$1.0 & $-$0.5 & 1.8 & 2.0 & 0.3 & 7.6 & $-$1.6 & 8.3 \\
 & Qwen3 32B & 3.2 & $-$0.7 & $-$0.7 & 1.3 & $-$0.4 & 1.3 & $-$2.2 & 10.7 & 5.3 & 3.8 \\
\bottomrule
\end{tabular}
\end{table*}

\subsection*{Individual-level skill appears only where profiles predict behavior}

\begin{figure*}[t]
\centering
\includegraphics[width=\textwidth]{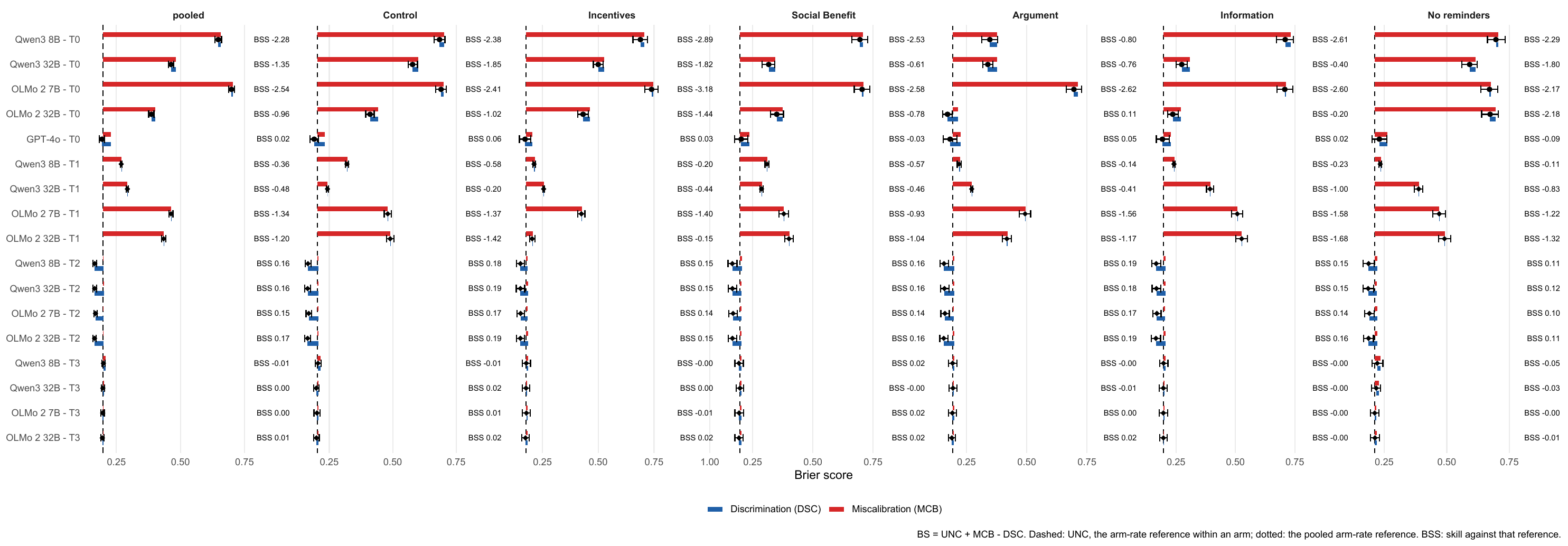}\\[2pt]
\includegraphics[width=0.75\textwidth]{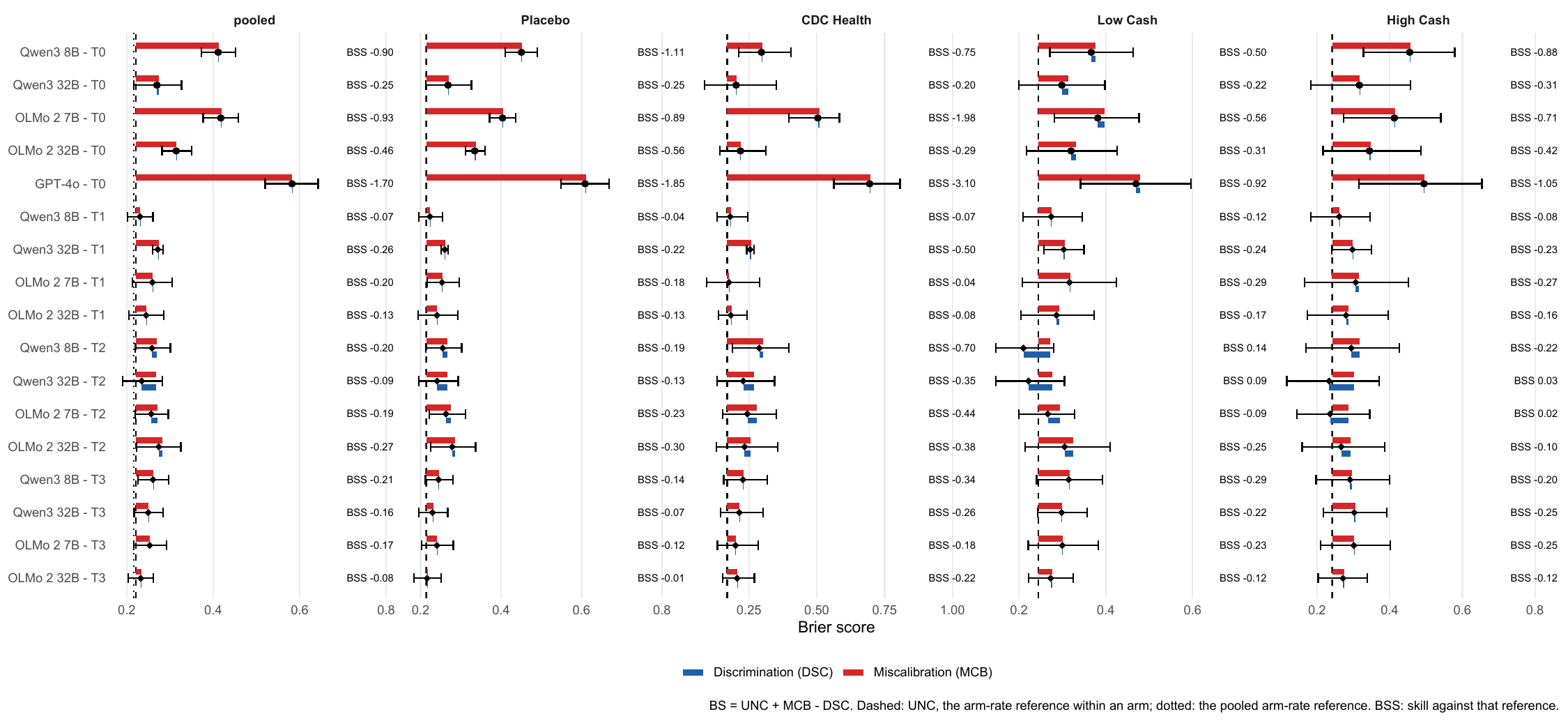}
\caption{\textbf{Individual-level accuracy (D3).} Brier score of the predicted probability of vaccination on the held-out units, pooled and by arm, by training protocol (rows), for (\textit{A}) Sweden and (\textit{B}) Ghana, with 95\% bootstrap intervals. Each score is built from its CORP decomposition, $BS = UNC + MCB - DSC$: the vertical line marks the irreducible uncertainty $UNC$ of the outcome, which equals the Brier score of a forecaster who predicts the arm rate; the bar to the right adds the miscalibration $MCB$, and the bar to the left subtracts the discrimination $DSC$. The Brier skill score against the arm-rate reference is printed in each row; zero means no information beyond the arm rate. Only pilot tuning in Sweden (T2) has substantial positive skill, equal to that of a logistic regression on the profile's four vaccine-attitude items.}
\label{fig:d3}
\end{figure*}

Matching an arm's vaccination rate is a different task from identifying which participants vaccinate. Meaningful individual-level skill appears only in Sweden after pilot-only tuning (T2). There all four models reduce the squared prediction error by 15--17\% relative to a forecast of the arm's rate (Brier skill scores of 0.15--0.17; Fig.~\ref{fig:d3}; SI Appendix, Table S3). Their probabilities are well calibrated (miscalibration below 0.004) but separate vaccinators from nonvaccinators only modestly (discrimination of 0.03--0.04). A logistic regression on the four vaccine-attitude items of the Swedish questionnaire, fitted to the same pilot, attains almost the same skill (0.17) and discrimination (0.036; exploratory; SI Appendix, section 9).

Elsewhere individual-level skill is negligible. In Sweden, training on the corpus before the pilot (T3) all but removes the advantage of T2 (skill scores of $-0.01$ to 0.01). In Ghana vaccination is strongly clustered by village (intraclass correlation 0.65), and the demographic profile does not identify which villages vaccinate. Individual accuracy also says little about the accuracy of treatment effects. A treatment effect of 5 pp changes the Brier score by only about 0.002, which is how the Swedish T2 models can predict individuals best while still missing the payment effect.

\subsection*{Precision gains are modest and matched by simple regression}

\begin{figure*}[t]
\centering
\includegraphics[width=\textwidth]{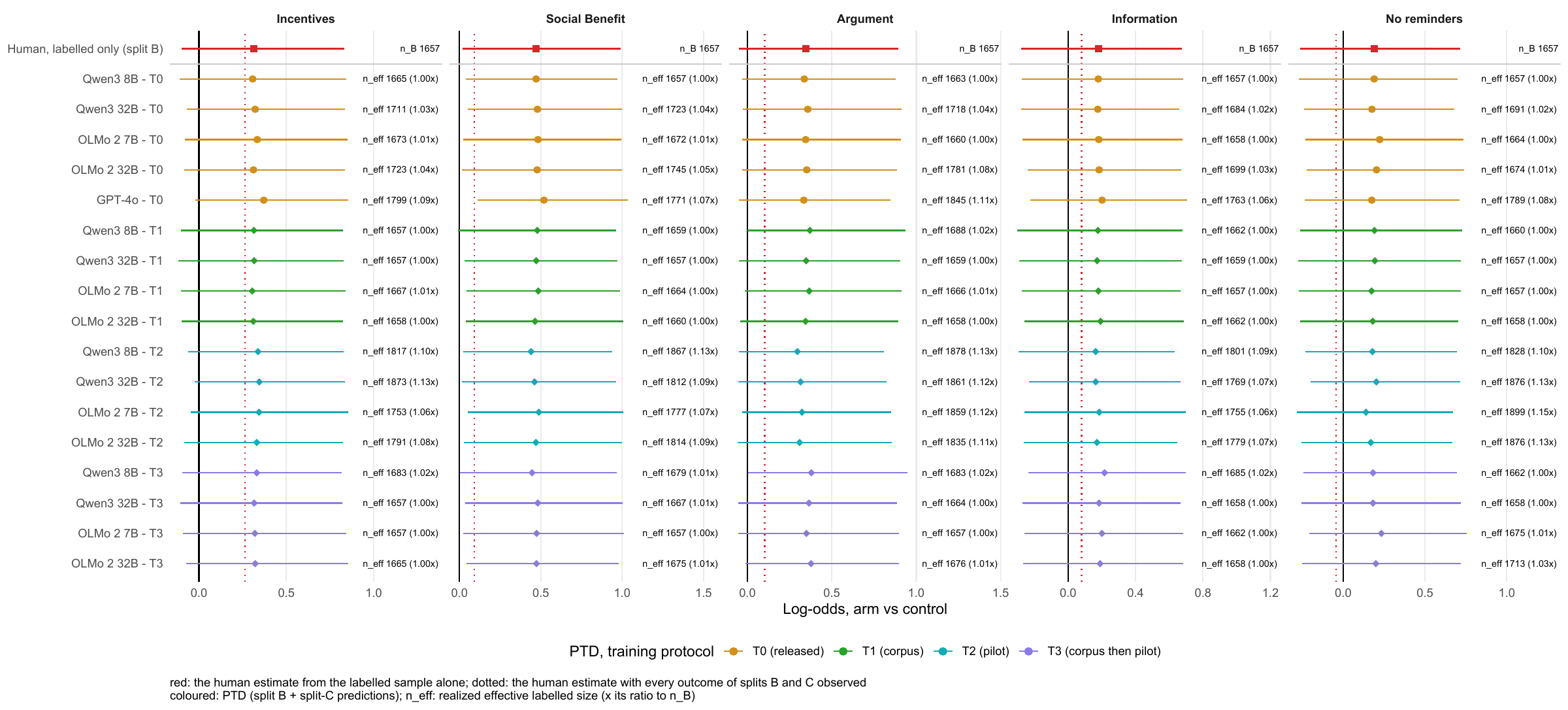}\\[2pt]
\includegraphics[width=0.75\textwidth]{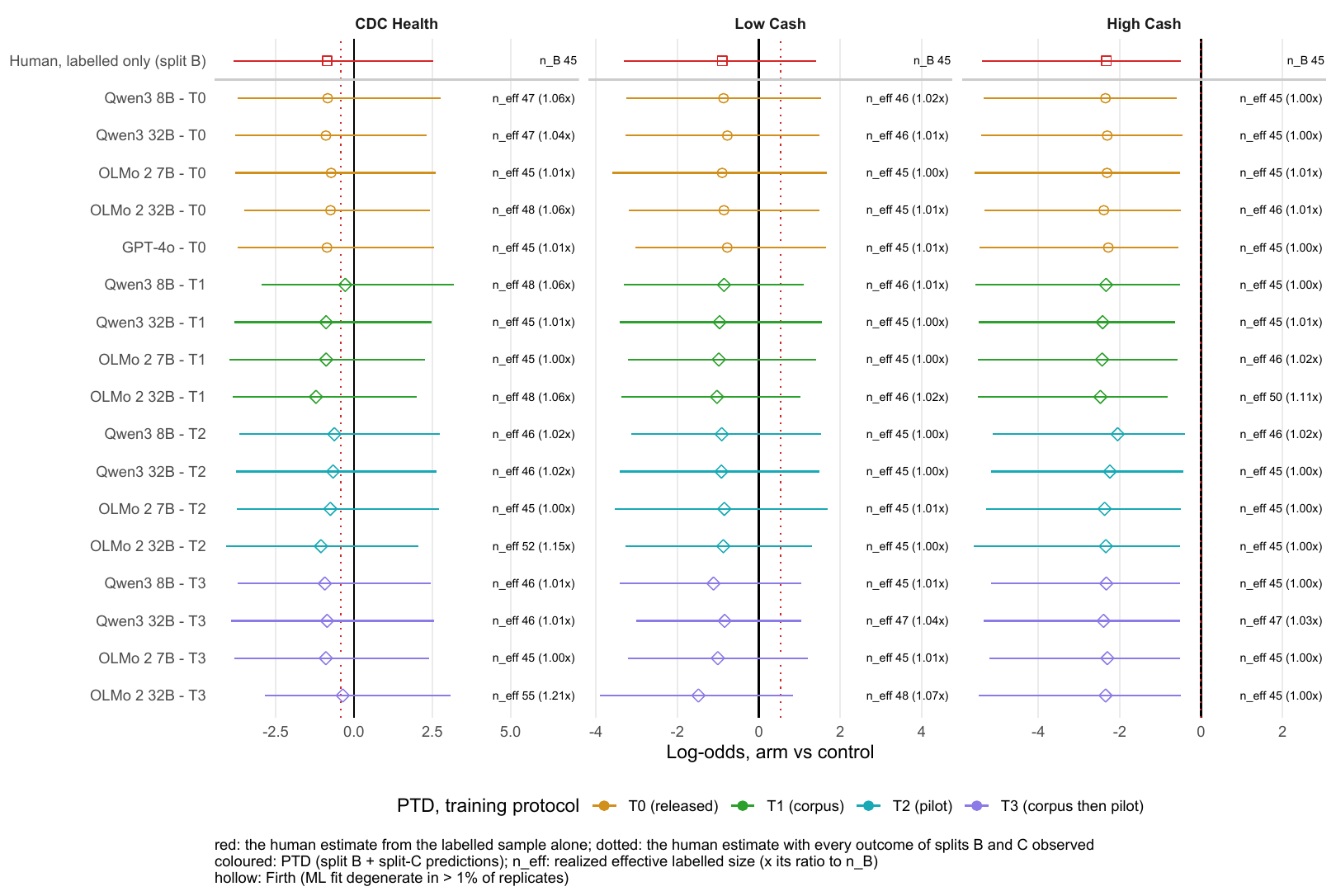}
\caption{\textbf{Precision gains under Predict-Then-Debias inference (D4).} For each active arm (panels), the top row shows the human estimate of the log-odds effect from the labeled sample alone (split B: 1,657 participants in Sweden, 45 villages in Ghana) with its 95\% percentile interval. The rows below show, for each training protocol, the PTD estimate that augments the labeled outcomes with the model's predictions on the held-out split C, with its 95\% interval from 2,000 bootstrap replicates (participants in Sweden, villages in Ghana), and the realized effective labeled sample size with its ratio to the labeled size. The dotted line is the human estimate with every outcome of splits B and C observed, the target that the augmented estimate approaches as predictions improve. Hollow markers are Firth-penalized labeled fits, reported where the maximum-likelihood fit separated in more than 1\% of replicates. (\textit{A}) Sweden: gains of up to 15\% under pilot-only tuning and 11\% for GPT-4o, and none above 3\% elsewhere. (\textit{B}) Ghana: no gain distinguishable from zero.}
\label{fig:d4}
\end{figure*}

For estimation, the question is whether synthetic predictions sharpen a trial's estimates when they are combined with a labeled human sample, however biased the predictions may be. The PTD estimator starts from the estimate implied by the predictions on the held-out sample, which is precise but possibly biased, and corrects it by the prediction error measured on the labeled sample. A tuning weight for each coefficient sets the size of the correction, so that informative predictions lend the precision of the larger held-out sample, while uninformative ones are down-weighted until the labeled-only estimator is recovered. The model need not be calibrated, let alone correct: poor predictions cost efficiency, but the correction protects the estimand. We express the gain as an effective labeled sample size, the number of labeled participants the human-only estimator would need to match the precision of the augmented one. In Sweden, where the labeled sample holds 1,657 participants, a gain of 15\% is worth about 250 additional participants. Fig.~\ref{fig:d4} reports the gain for every arm and training protocol, together with the quantity that governs it \citep{Broskaetal2025}. Each participant in the labeled sample (split B) contributes to the estimate twice over: once through the outcome actually observed, and once through what the predicted probability would contribute in its place. The correlation between these two contributions across labeled participants is the PPI correlation, $\tilde\rho$. For a simple mean it reduces to the plain correlation between outcomes and predictions; for the trial's regression it is the same idea once arm and covariates are accounted for.

In Sweden the realized gains range from 0 to 15\%. They are concentrated where individual-level skill was found: 6--15\% under pilot-only tuning and 6--11\% for GPT-4o, whose miscalibrated probabilities nonetheless rank participants ($\tilde\rho$ up to 0.36), against 0--3\% under corpus-then-pilot tuning and 0--2\% under corpus tuning. The gain is predictable, and it does not depend on what produces the predictions. Across the 85 Swedish combinations of arm, model and training protocol, the realized gain is a function of $\tilde\rho$ alone. It is predicted almost exactly by the closed-form expression $1/(1 - 0.75\,\tilde\rho^{\,2})$, in which 0.75 is simply the held-out sample's share of the evaluation data ($60/(60+20)$): the gain the formula predicts and the gain realized differ by 0.01 on average -- a predicted 1.10 against a realized 1.09, say (SI Appendix, section 9). A source of predictions is therefore worth exactly the correlation it attains. The four-item logistic regression of the previous section, with arm indicators added and used as the source of predictions, attains correlations similar to the models' ($\tilde\rho$ = 0.30--0.43 across arms, against 0.29--0.39 for the pilot-only models) and gains of 9--15\%, within 3 pp of the best model's gain in every arm (SI Appendix, section 9).

What the estimator exploits is the graded probability, with its full ordering of participants, not a realized Yes/No answer drawn from it. When it is given, in place of each participant's predicted probability, a Yes or No drawn from that probability, which is a synthetic respondent in the literal sense, most of the gain disappears. For the pilot-only models the mean effective-sample ratio across arms falls from 1.10 to 1.02, and for GPT-4o, whose probabilities lie close to 0 or 1 so that a draw loses little, it falls from 1.08 to 1.06 (SI Appendix, section 9). The draw adds noise that erases the ranking the probabilities carry. Calibration matters just as little, because the correction step absorbs systematic error in the level of the probabilities. This is how the miscalibrated probabilities of GPT-4o still earn a gain: they rank participants correctly even where their levels are wrong.

In Ghana, where the labeled sample consists of 45 villages, no gain can be distinguished from zero. The labeled sample is at the limit of what the estimator can use. Its high-cash arm contains 3 vaccinated participants out of 81, so that when the bootstrap resamples whole villages, 37--40\% of the replicates contain an arm in which every outcome is identical and the logistic regression cannot be estimated. We therefore report the Ghanaian estimates with a penalized estimator that remains defined in such replicates (\textit{Materials and Methods}). The penalized estimates show gains of up to 21\%, but these gains carry no evidential weight: the correlation from which they derive is estimated with 95\% CIs of mean width 0.58, and the interval excludes zero in only 1 of the 51 combinations of arm, model and training protocol (SI Appendix, Fig.~S5). % s6-main-results-begin
Two further checks reach the same conclusion by different routes. The wild cluster bootstrap, the check registered for designs with few treated clusters, reproduces the width ratios on which the result rests to within 0.09 (SI Appendix, Table S4). In the coverage simulation, however, its own intervals contain the true coefficient in only 77--88\% of simulated datasets, against 90--97\% for the prediction-powered cluster-bootstrap interval, so the cluster-bootstrap intervals are the ones we report (SI Appendix, section 7). A closed-form cluster-robust estimator \citep{BroskaHowes2026}, which chooses how much weight to place on the predictions and confines that weight between 0 and 1, finds almost nothing worth weighting: the weight is zero in 31 of the 51 Ghanaian arms and never exceeds 0.13, while in Sweden its gains track those of the registered estimator (SI Appendix, section 9). The OLMo~2 models in Ghana are the primary test of augmentation in our design, because OLMo~2's training data demonstrably exclude the Ghanaian trial (SI Appendix, section 8), and they are among the cases in which augmentation yields nothing.

\subsection*{What each training protocol teaches}
To learn what a model takes from its training data, we compare it with a logistic regression fitted to the same data that reads the same participant profile but involves no language model (\textit{Materials and Methods}; SI Appendix, section 9). A model that predicts new participants no better than this regression has learned nothing that its training data did not already contain in a form a regression can extract.

The corpus supplies a level but not knowledge. In both trials the corpus-tuned models place the arms near the corpus's pooled vaccination rate of 30\%: the two OLMo~2 models within 3--9 pp of it and 31--46 pp from the rate of the corpus sources from the same country, so the country named in the prompt does not move the level. They do not learn how vaccination varies with the participant's characteristics in the corpus. A regression fitted to the corpus and applied to the trial participants -- the map a statistician would extract from the same training data -- explains at most 5\% of the variation in their predictions in Sweden, and in Ghana it orders participants in the opposite direction to the models. What moves their predictions is the wording of the treatment, which alone accounts for 61--72\% of the variation in the predictions of three of the four models, against 0.2\% of the variation in the human outcome. That sensitivity, anchored neither to levels nor to people, is the source of the erratic corpus-tuned effects, such as the range from $-5$ to 45 pp assigned to the same Swedish payment. The corpus could not even have taught that cash raises uptake. Of its two cash incentives paired with an observed behavior, one raises uptake by 1 pp and the other reads 14 pp below its control because of a labeling error carried by every reported run (SI Appendix, section 3); the remaining money treatments shift stated intentions by a few percentage points at most. The average response of the corpus-tuned models to the Swedish payment, an increase of 17 pp, amplifies a response the released models already show, which places its origin in pretraining.

The pilot is learned through whatever in the profile best predicts its outcomes, and what is learned travels only if that information describes people. In Sweden the most predictive items are the participants' attitudes to vaccination, which hold outside the pilot. The pilot-only models learn them, and a regression fitted to the pilot reproduces about four-fifths of what they predict. Their arm rates are therefore the pilot's, sampling error included, and their ranking of participants is the regression's: the four attitude items give them the skill of a four-item regression and the whole of their precision gain. In Ghana, where vaccination is decided village by village, the most predictive information in the prompt is the name of the village, and the pilot-only models learn which pilot villages vaccinated. Village indicators explain 96--99\% of the variation in their predictions, and the village rates they predict barely match those of villages they never saw (correlation 0.17--0.45, and at most 0.18 within a district). What they learned is local. Outside the pilot it reproduces the district rates for which the trial's regression already adjusts, which is why it passes the noise floor without yielding a gain in precision. In neither trial do the pilot-only models outperform the map a statistician would extract from the same training data: a regression fitted to the pilot and applied to the held-out participants matches them on treatment effects, on arm rates and on individual outcomes alike (SI Appendix, section 9).

Training on the corpus and then on the pilot does not add the two sources; the corpus dilutes what the pilot teaches. A regression fitted to the pilot and applied to the held-out participants can reproduce the pilot-only models' predictions almost person for person, matching about four-fifths of their variation across participants; of the corpus-then-pilot predictions it reproduces only 9--21\%: the pilot's map is only faintly imprinted on the corpus-trained model, and the gains the pilot brings at the individual level disappear with it (Brier skill 0.00 against 0.16; precision gains at most 3\% against 15\%). In Ghana the corpus-then-pilot effects are the smallest of any training protocol, but they are no closer to the human effects than those of a regression combining the corpus and the pilot, so nothing in their smaller errors can be credited to what the models learned.

Across training protocols the conclusion is the same. In 72 comparisons between a fine-tuned model and the regression fitted to its training data, on treatment effects, arm rates and individual predictions in the held-out sample, the model is more accurate once, less accurate in 26 and indistinguishable in the rest (SI Appendix, section 9). The same holds for precision. A regression fitted to the Swedish pilot raises the effective labeled sample by 7--19\%, against 6--15\% for the pilot-only models, and in no arm does the best of these models exceed it by more than 2 percentage points; the regression combining corpus and pilot gains 10--15\%, against at most 3\% for the corpus-then-pilot models. In Ghana no regression, like no model, achieves a gain distinguishable from zero. The treatment effect resides in none of the training sources: the pilot carries it only together with sampling noise of the same order, the profile does not contain it, and the corpus holds it for other interventions in other populations, one contrast per study, which the model absorbs as a single pooled rate.

\begin{table*}[t]
\centering
\caption{\textbf{Summary by training protocol.} Entries pool the four open-weight models (GPT-4o is shown separately); ranges are across models, and across arms for D2. D1: share of model--arm comparisons in which the synthetic effect has the sign of the human effect, and root-mean-square error (RMSE) of the synthetic effects over all arms of the trial, every arm weighted equally (exploratory). Human sampling noise alone produces an RMSE of up to 3.5 pp in Sweden and 14.3 pp in Ghana (97.5th percentiles over 2,000 bootstrap resamples of the held-out sample). All T2 and T3 arms lie inside the preregistered joint 95\% prediction interval. D2: absolute error in the arm vaccination rate and share of arms below the preregistered noise floor. D3: Brier skill score against the arm rate, pooled over arms. D4: realized effective labeled sample size relative to the labeled sample. The last rows score benchmarks that use no language model: logistic regressions fitted to the pilot, to the corpus and to both, which read the same profile as the models (SI Appendix, section 9; their D1 entries come from the fractional logit of the registered specification), and a forecast of no effect in every arm. Full results by training protocol and arm are in SI Appendix, Tables S1--S4.}
\label{tab:summary}
\small
\begin{tabular}{llcccccc}
\toprule
 & & \multicolumn{2}{c}{D1 effects} & \multicolumn{2}{c}{D2 rates} & D3 skill & D4 gain \\
\cmidrule(lr){3-4}\cmidrule(lr){5-6}
Trial & Training protocol & sign (\%) & RMSE (pp) & error (pp) & below floor (\%) & (pooled) & $n_{\mathrm{eff}}/n_B$ \\
\midrule
Sweden & T0 released, open-weight & 85 & 2.3--24.7 & 6.9--75 & 0 & $-2.5$ to $-1.0$ & 1.00--1.08 \\
       & T0 released, GPT-4o & 80 & 2.9 & 10--16 & 0 & $+0.02$ & 1.06--1.11 \\
       & T1 external corpus & 65 & 8.0--19.1 & 13--57 & 0 & $-1.3$ to $-0.4$ & 1.00--1.02 \\
       & T2 pilot & 40 & 2.3--3.0 & 0.6--6.6 & 83 & $+0.15$ to $+0.17$ & 1.06--1.15 \\
       & T3 corpus then pilot & 45 & 1.3--3.3 & 0.2--6.1 & 83 & $-0.01$ to $+0.01$ & 1.00--1.03 \\
\addlinespace
       & Pilot regression (no model) & 40 & 2.1 & 0.3--4.3 & 83 & $+0.18$ & 1.07--1.19 \\
       & Corpus regression (no model) & 60 & 4.0 & 20--26 & 0 & $-0.35$ & 1.00--1.02 \\
       & Combined regression (no model) & 40 & 2.3 & 0.3--4.6 & 83 & $+0.18$ & 1.10--1.15 \\
       & No effect (no model) & -- & 2.3 & -- & -- & -- & -- \\
\addlinespace
Ghana  & T0 released, open-weight & 58 & 2.0--44.5 & 0.6--57 & 19 & $-0.9$ to $-0.2$ & 1.00--1.06 \\
       & T0 released, GPT-4o & 33 & 14.0 & 47--72 & 0 & $-1.7$ & 1.00--1.01 \\
       & T1 external corpus & 58 & 1.1--22.1 & 0.3--28 & 56 & $-0.3$ to $-0.1$ & 1.00--1.11 \\
       & T2 pilot & 58 & 6.1--13.3 & 2.0--28 & 69 & $-0.3$ to $-0.1$ & 1.00--1.15 \\
       & T3 corpus then pilot & 83 & 3.8--8.3 & 2.2--13 & 100 & $-0.2$ to $-0.1$ & 1.00--1.21 \\
\addlinespace
       & Pilot regression (no model) & 33 & 16.4 & 4.4--30 & 75 & $-0.03$ & 1.00--1.03 \\
       & Corpus regression (no model) & 67 & 10.2 & 5.2--14 & 100 & $-0.18$ & 1.00--1.03 \\
       & Combined regression (no model) & 33 & 11.1 & 4.7--22 & 100 & $-0.02$ & 1.01--1.04 \\
       & No effect (no model) & -- & 10.2 & -- & -- & -- & -- \\
\bottomrule
\end{tabular}
\\[2pt]
{\footnotesize Ghana D4 from the Firth-penalized labeled fits; no Ghanaian gain is distinguishable from zero (the interval of $\tilde\rho$ excludes zero in 1 of 51 combinations of arm, model and training protocol). The T0 open-weight rows exclude GPT-4o. The smallest Ghanaian RMSE of T0 and T1 (2.0 and 1.1 pp) belong to Qwen3 32B, for which exposure to the published results cannot be excluded.}
\end{table*}

\section*{Discussion}

We asked what an experimentalist gains from synthetic participants in a single trial, both as a forecast of its results and as an input to its estimation, and which of the data at the experimentalist's disposal produce the gain. Specializing a released model to the trial, by fine-tuning it on related studies and on a pilot, yields synthetic samples that resemble the human sample but do not recover what the trial was run to measure. Fine-tuning on the pilot brings the vaccination rate of most arms within sampling error of the human rate and, in one of the two trials, allows the model to rank participants by their probability of vaccinating. It does not recover the treatment effects. The synthetic contrasts follow the pilot's, and across all eight contrasts their errors cannot be told apart from those of forecasts that use no model. The gain in precision is at most 15\% of the labeled sample, and a regression on four questionnaire items matches it in every arm. Models that have not seen the trial's data misstate its vaccination rates by wide margins and, apart from two cases in which exposure to the published results cannot be excluded, its treatment effects as well.

The four evaluation dimensions do not agree with one another, and their disagreement is informative. Pilot-tuned models produce predictions that are hard to tell apart from the human data: the share vaccinating in each arm lies within sampling error of the observed share in most arms of both trials, and in Sweden the probability assigned to each participant matches the frequency with which similar participants actually vaccinated. Yet the same samples miss what the trials were run to measure, because their treatment effects follow the pilot's contrasts rather than those of the held-out sample. The realism and the failure have a common source. The models reproduce the pilot, so the synthetic vaccination rates inherit its sampling error and the synthetic effects inherit its noise. Resemblance to the human data is therefore not evidence of validity. The same divergence between resemblance and accuracy in effects has been documented across many online experiments \citep{Hewittetal2024, Cuietal2025}, and recent work argues that it holds as a general principle: behavioral fidelity is neither necessary nor sufficient for valid statistical substitution \citep{Wangetal2026}, and matched answer distributions can coexist with synthetic respondents unlike any human \citep{KumagaeKaneko2026}. However a limited human-data budget is allocated across prompting, fine-tuning and post-hoc rectification, it is the rectification step, not the resemblance of the generations, that secures valid estimates \citep{Krsteskietal2025}. The single-trial design adds the mechanism behind the divergence, the copying of the pilot, and with it a practical rule: synthetic participants must be validated on the effect they are meant to forecast, not on their resemblance to the people they imitate. 

The lesson is familiar elsewhere: accuracy measured on cases that overlap with the training data overstates accuracy on genuinely new cases. In clinical prediction the overlap is total -- a model's fit to the patients it was derived on overstates its accuracy on new patients \citep{Steyerbergetal2001}. In molecular biology it is partial -- predictors of protein interactions look far better on test pairs that share a protein with the training pairs than on pairs that share none \citep{ParkMarcotte2012, Bernettetal2024}. In our design the shared component is the village name: a model that has memorized each pilot village's vaccination rate under its name looks accurate wherever that name reappears, without having learned anything about behavior.

Our results add a second caution, about the yardstick rather than the model. In a trial of ordinary size the human effect estimate is itself noisy, so agreement with it cannot validate a forecast of effects on its own; a forecast should also be compared with what the pilot's raw contrast, or no forecast at all, would have delivered -- and in our two trials the pilot-tuned forecasts do no better than either.

The results also bound what augmentation can deliver, and the bound is set by the data rather than by the model. Under PTD the gain depends on a single quantity: how well the predicted probabilities track who actually vaccinates among the held-out participants, once the trial's own regression has adjusted for the covariates it already uses. Only prediction beyond those covariates buys precision, and the realized gains follow this residual correlation closely. The correlation, in turn, is a property of the information the prompt carries. Trials and surveys record their participants as tables of covariates, and a profile rendered from such a table carries two kinds of information: the covariates for which the trial's regression already adjusts, which contribute nothing further, and whatever predictive items the questionnaire happened to include. Where the Swedish profile contained four attitude items, a gain appeared, in a released model as much as in a pilot-tuned one, and a short regression on the same items reproduced it. Where the Ghanaian outcome was a village event and the profile demographic, nothing any model did could produce one. The size of the augmentation gain is best judged against what ordinary design choices deliver. In a cluster-randomized trial every resident of a village shares the village's treatment, so the effect can be estimated only by comparing villages with one another: each village is, in effect, a single observation. The registered design concentrates the labeled sample in 45 whole villages, leaving the other 137 of the 182 without a single measured outcome. Spreading the same number of labeled participants across every village instead would have brought most of them into the comparison: in our emulation the labeled participants span 134 villages, and the standard error of the human estimate falls by 31--41\% (39\% at the median), as much as labeling 2.7 times as many participants in whole villages would achieve. Under neither design is the gain from the synthetic predictions distinguishable from zero (SI Appendix, section 9). A reallocation of the same measurement budget is worth an order of magnitude more than the model. 

For the treatment effect itself, the constraint is not a bound but an absence: the effect is not in a covariate table at all. A pilot carries it only together with sampling noise of the same order as the effect, and a corpus of other trials carries one contrast per study, buried in variation across profiles. This ceiling, whereby a language model reading a covariate table does not outpredict a conventional statistical model fitted to the same table, is not peculiar to our trials. On tabular household panels a gradient-boosted tree outpredicts every one of thirteen LLMs \citep{Huangetal2026}; prompted personas fail to beat demographic baselines fitted to the same data \citep{Chenetal2026}; and persona conditioning improves simulation only when it is induced from in-domain behavioral data, the prompting analogue of our pilot \citep{Leeetal2026}. The pattern recurs well beyond social science. Machine learning shows no average gain over logistic regression in clinical prediction \citep{Christodoulouetal2019}, and single-cell foundation models pretrained on millions of cells do not outperform simple baselines on new tasks without task-specific training \citep{Kedzierskaetal2025, Boiarskyetal2024}. Where fine-tuning does succeed, the training data resemble the target closely: gene-network models fine-tuned on small samples of the task at hand \citep{Theodorisetal2023}, opinion models trained on surveys of the same population and question formats \citep{Suhetal2025}, and behavioral models scored on new participants of the tasks they were trained on \citep{Binzetal2025, OhGobet2026}. Our corpus lacked that resemblance to the target trials, and our pilot lacked size. Even the Swedish gain, at most 15\% of the labeled sample, sits at the lower end of the 10--30\% reductions in sample size that prognostic scores trained on large historical datasets deliver in clinical trials \citep{Schuleretal2022}.

What a language model can do and a regression cannot is read text. The route to larger gains therefore runs through richer, unstructured inputs: open-ended baseline answers, in-depth interviews, ethnographic and other qualitative records of the population and the setting, and descriptions of the clusters where randomization is clustered. Such sources carry the intentions, constraints and local circumstances that decide behavior and that a covariate table flattens, and they are the inputs on which a model's capacity to generalize from language is worth something. Richer profiles address who responds, however, not what a treatment does to behavior. Teaching the latter requires a different kind of corpus, organized around effects rather than individuals, in which interventions are described together with their doses, settings and measured outcomes, in the terms the model reads at inference, so that training can tie the description of a treatment to the direction and size of what it produced. Nor are richer inputs automatically better. Added profile detail has
failed to improve simulation where the added fields carried no predictive
signal \citep{Kohetal2026}; what matters is information, not length.
Early efforts to locate that signal converge: reorganizing the same
survey history into structured profiles helps only through its preference
content \citep{Yeetal2026}, and a person's own earlier choices outpredict
demographic, personality, and cognitive descriptions of who they are
\citep{PourtaheriZareei2026} --- the information that predicts a decision
is evidence about choices, not descriptions of the chooser. Both routes --- richer inputs about the person, and a corpus organized
around effects --- can be scored with the evaluation framework developed
here: the four dimensions apply unchanged to a model trained on
interviews or on an effect-centered corpus, and would show whether the
ranking sharpens and whether the effects stop being copies of a pilot.

Our results are subject to four qualifications. First, the study covers two trials and one behavioral domain, vaccination under incentives and nudges. It offers the depth of individual-level predictions of verified behavior, but neither the breadth of a megastudy nor that of an archive of effects. Second, the human benchmarks are themselves imprecise, so treatment-effect recovery, the primary dimension, cannot at this scale distinguish a damped effect from a recovered one in any single training protocol. What identifies the copying is not any single estimate but the dependence of the synthetic effects on the pilot's contrasts across all eight contrasts, together with the comparison against forecasts that use no model. The Ghanaian benchmark also depends on the sample. On all held-out units, in a regression without the social-network metric, the low-cash effect is 9.6 pp and not distinguishable from zero, and the one fully concordant combination loses that status (SI Appendix, section 9). Third, the two most striking individual successes, GPT-4o's reproduction of the published Swedish estimate and the full concordance of Qwen3 32B in Ghana, are the two cases in which exposure to the published papers cannot be excluded. The released training data of OLMo~2 contain the Swedish paper but not the Ghanaian one, which is why the Ghanaian OLMo~2 models are the primary case for augmentation, and that is the case in which no gain is detectable. Finally, every model enters the design in its instruction-tuned form, as providers release them. There is growing evidence that alignment for assistant use itself works against human simulation, by compressing the diversity of responses that a population of real participants exhibits \citep{Kangetal2026}; a base model is therefore a lever our ablation did not vary. Prospective validation, with synthetic predictions registered before a trial is run and with models whose training data are known, is the test the approach now needs.

What synthetic participants can do for an experimentalist today is modest and specific. They reproduce a pilot's rates, which has uses in the design of instruments and in power calculations but adds no information to what the pilot already contains. They rank participants where the profile allows it, and the debiased estimator turns that ranking into a gain in precision that is small, predictable in advance from the correlation, and no larger than that of conventional modeling of the same items. They do not supply the one quantity the trial is run to learn, the size of the treatment's effect. Progress will come from inputs richer than the covariate table, and from validation on effects rather than on resemblance.

\matmethods{

\subsection*{Target Trials.}
The Ghanaian trial \citep{Duchetal2023} is a cluster-randomized trial across villages in six rural districts of Ghana's Central Region. Each village was assigned to one of four 45-second videos delivered in February 2022: a placebo, a standard health message, a low cash incentive (20 GHS, about \$3) or a high cash incentive (60 GHS, about \$10), with the incentives conditional on vaccination within six weeks. The outcome is vaccination by April 2022, verified against District Health Office records; the analytic sample comprises the 3,075 participants in 234 villages whose status could be verified. The Swedish trial \citep{CamposMercadeetal2021} is an individually randomized trial of 8,286 Swedish adults aged 18--49 from an online panel, assigned between May and July 2021 to a control condition or to one of five treatments: a guaranteed payment of SEK~200 (about \$24) conditional on vaccination, three behavioral nudges (an information quiz, an active-arguments task and a social-impact prompt) and a no-reminders condition. The outcome is first-dose vaccination within 30 days of eligibility, recorded in national administrative registers. In both trials we estimate the treatment effects on the log-odds scale by logistic regression with the original study's covariates. For Ghana these are gender, centered age, education, centered income, employment, the centered social-network metric, centered distance to the clinic and district fixed effects, with standard errors clustered by village; the placebo arm pools the placebo villages with the embedded placebo in treated villages, as in the original analysis. For Sweden they are gender, age band by region fixed effects, COVID-19 risk group, civil status, children, employment, education, parents' birthplace and income, with heteroskedasticity-robust standard errors. The Ghanaian regressions use complete cases on these covariates, as the original did (1,350 of 1,929 held-out units; SI Appendix, section 2). On all 1,926 held-out units, in a regression without the social-network metric, the low-cash effect is 9.6 pp (OR 1.66, 95\% CI 0.85--3.23) and not distinguishable from zero; the registered comparison is the complete-case one, and the sensitivity analysis is reported in SI Appendix, section 9.

\subsection*{Sample Splits.}
Each trial is partitioned into three disjoint splits, stratified by arm, at the village level in Ghana and at the participant level in Sweden. Split A (20\%) is the pilot on which the two pilot-tuning protocols (T2 and T3) are trained; split B (20\%) is the labeled sample for the prediction-powered estimator; and split C (60\%) is the held-out sample whose outcomes score the first three dimensions (D1--D3) and are treated as unobserved in the fourth (D4). Because the splits are disjoint, no unit is both trained on and later used to evaluate the model or to correct its predictions. Splits A and B are excluded from the evaluation in every training protocol, so the evaluation sample is identical across training protocols. For Ghana, split A contains 623 participants, split B 373 complete-case participants in 45 villages, and split C 1,929 participants (1,350 complete cases in 137 villages); for Sweden, the three splits contain 1,657, 1,657 and 4,972 participants.

\subsection*{Models and Training Protocols.}
Four open-weight instruction-tuned models, OLMo~2 7B and 32B \citep{OLMo2} and Qwen3 8B and 32B \citep{Qwen3}, enter all four training protocols. GPT-4o (\texttt{gpt-4o-2024-05-13}) \citep{GPT4o} is included for comparison with a model family widely used in the literature on synthetic participants. It enters T0 only, because its fine-tuning interface does not expose the hyperparameters that reproducible fine-tuning requires \citep{Spirling2023}. OLMo~2 releases its training data, which allows a direct test of exposure to the target trials (see \textit{Exposure of the Training Data to the Target Trials}). The first protocol (T0) uses the model as published. The corpus protocol (T1) fine-tunes it on the external corpus, and the pilot protocol (T2) fine-tunes it on the pilot. The corpus-then-pilot protocol (T3) starts from the T1 adapter and continues training on the pilot alone, as registered, so that it measures the marginal contribution of the pilot on top of the corpus rather than that of a jointly trained mixture.

\subsection*{External Corpus.}
The corpus pools eight published experiments and six surveys that record an individual-level vaccination outcome (verified, self-reported or intended), report at least age and gender for every unit and provide microdata. They were chosen for their proximity to the target trials in outcome, setting and intervention and for their coverage of both income tiers, and they contribute 69,490 experimental and 18,190 survey records, 87,680 in all (SI Appendix, section 3 lists the sources). The composition and size are those registered in the plan, and the target trials never enter the corpus.
From each source we build a unit-level table with the covariates, one outcome and, where present, the treatment indicator. We exclude administrative fields, measurements taken after treatment that the treatment could plausibly have altered, items conditional on the outcome and items without fixed wording. Each covariate is rendered as a question--answer pair in the source's own wording, with the answer set embedded where the values are not self-explanatory. Prompts are capped at 4,096 tokens. One source, the Ghanaian newborn-vaccination trial, carries a labeling error found after the runs: children born before the intervention are rendered with their community's treatment transcript. Every reported run used the corpus as it stands (SI Appendix, section 3).

\subsection*{Prompting and Inference.}
Every unit is rendered through one schema (SI Appendix, section 4): a system message that instructs the model to simulate a human with the given profile, followed by the profile as question--answer turns, the study context and the assigned treatment text; and a user message that poses the outcome question and restricts the answer to the source's label set. At training time the observed outcome is the assistant turn, and the loss covers it alone. At inference each unit is queried once, at temperature 1.0 and without truncation \citep{LeeYue2026}, and the main analyses use the predictions for the units held out from training. The synthetic outcome $\hat p_i$ is the probability of Yes renormalized against the probability of No at the first position that answers Yes or No, read from the five most probable tokens. The renormalization removes the probability mass spread across every other string (formatting tokens, hedges, near-synonyms), which would otherwise deflate and distort a raw token probability \citep{Holtzmanetal2021}, and puts every model on a common $[0,1]$ scale however its tokenizer spells Yes and No. The realized Yes/No response is recorded but enters no metric. Qwen3's reasoning mode is disabled, so that the first generated tokens are the answer itself. Units lacking either token at the read position are excluded (at most 3 per combination; SI Appendix, Table S5).

\subsection*{Fine-Tuning.}
We use low-rank adaptation \citep{Huetal2022} with rank 32, $\alpha = 64$ and dropout 0.05 on the attention and feed-forward projections, trained with the TRL supervised fine-tuning trainer on AWS SageMaker for six epochs with a cosine schedule, a warm-up ratio of 0.03, weight decay of 0.01, the AdamW optimizer, bf16 precision and seed 42. The effective batch size is 8 and the learning rate $2\times10^{-4}$ for the pilot corpora, and 64 and $5.66\times10^{-4}$ for the external corpus, the learning rate scaling with the square root of the batch size, as registered. No validation set or early stopping is used; the adapter from the final epoch is kept and applied unmerged at inference. Full hyperparameters, instances and compute are reported in SI Appendix, section 5, together with the training-loss curves of T1--T3 that the plan specifies as a manipulation check.

\subsection*{Treatment-Effect Recovery (D1).}
For each combination we fit the trial's regression to binary outcomes drawn from the predicted probabilities, $\hat y_i \sim \mathrm{Bernoulli}(\hat p_i)$, and take the mean coefficient over $M = 2{,}000$ draws. Its 95\% interval comes from $B = 2{,}000$ bootstrap replicates that resample evaluation units (villages in Ghana) and redraw the outcomes, so that it reflects variation both across units and across outcome draws. Drawing the outcome, rather than fitting the regression to $\hat p_i$ directly, gives the synthetic estimate the sampling variation of a binary outcome; without it, significance concordance would compare an almost noiseless synthetic estimate with a noisy human one. Only the response differs from the human regression, so the synthetic coefficient is a like-for-like contrast on the same covariates. The human regression is fitted on the same units with the observed outcomes and design-based standard errors, which are those of the original analyses: the heteroskedasticity-robust sandwich variance scaled by $N/(N-1)$ in Sweden, and the cluster-robust sandwich variance over villages scaled by $G/(G-1)$ in Ghana.

Following the multi-indicator practice of the reproducibility literature \citep{OSC2015, Camerer2018}, we report four indicators for each arm. Sign concordance records whether the synthetic and human effects have the same sign. Significance concordance records whether they also agree on being distinguishable from zero (by a Wald test at the 5\% level for the human effect, and by the percentile interval for the synthetic one). The discrepancy between the two effects is reported on the log-odds scale and as an average marginal effect in percentage points, each with a joint-bootstrap interval. Finally, because a test against the null can return discordance from sampling variation alone even when the synthetic effect is consistent with its human counterpart, we record whether the synthetic effect lies inside a 95\% prediction interval for a replication \citep{Patiletal2016}, the range of synthetic effects to be expected once the human effect has been observed. The interval is formed from the joint bootstrap by centering the replicate differences on the observed discrepancy, and also in its normal-approximation form. A fixed-effect cell with no variation in the outcome in a given draw is dropped with its units before the fit, which leaves the arm coefficients unchanged. An arm indicator whose drawn outcomes are all alike is retained and flagged; where more than 1\% of fits are affected, Firth-penalized estimates are reported alongside. SI Appendix, section 2 records the departures from the plan's text on these points.

As an exploratory complement, we summarize the error of each combination over all contrasts of a trial by the root-mean-square difference between the synthetic and human average marginal effects, with every contrast weighted equally. Two reference values put this error in scale. The first is a noise floor analogous to that of the second dimension: the 97.5th percentile of the same statistic when the human regression is refitted on 2,000 bootstrap resamples of the held-out sample (participants in Sweden, villages in Ghana), which approximates the error that a forecaster who knew the true effects would still incur against this benchmark. The second is the error of two forecasts that use no model: the original specification fitted to the pilot alone, and a forecast of no effect in every arm.

\subsection*{Distributional Divergence (D2).}
For each arm we compute the Hellinger distance between the Bernoulli distributions defined by the human rate and by the mean predicted probability, the signed error in the rate and, for each combination, the mean distance weighted by sample size. The noise floor is the 97.5th percentile of the distance between the full-sample human rate and the rate in $B = 2{,}000$ resamples of the human labels alone (villages in Ghana), following the construction of ref.~\citep{Kollurietal2025}; a distance below the floor is read as no larger than sampling noise. The same resamples give a one-sided exceedance $P$ value.

\subsection*{Individual-Level Accuracy (D3).}
We compute the Brier score of $\hat p_i$ against the realized outcome \citep{Brier1950}, pooled and by arm, and the Brier skill score against the arm-rate reference, together with the CORP decomposition into miscalibration, discrimination and uncertainty \citep{Dimitriadisetal2021}, with bootstrap intervals (villages in Ghana).

\subsection*{Prediction-Powered Inference (D4).}
The PTD estimator \citep{Klugeretal2025}, $\hat\theta = \hat\theta(B, y) + \Omega[\hat\gamma(C, \hat p) - \hat\gamma(B, \hat p)]$, combines the labeled-only estimate on split B with the difference between the prediction-based estimates on splits C and B, the prediction-based fits taking $\hat p_i$ as a fractional outcome. The diagonal tuning matrix $\Omega$ minimizes the asymptotic variance and is estimated once and held fixed. Intervals are percentile intervals over $B = 2{,}000$ replicates that resample participants (Sweden) or villages (Ghana) with their labeled or unlabeled role intact (the cluster bootstrap, Algorithm 5 in ref.~\citep{Klugeretal2025}). We report the ratio of the width of the PTD interval to that of the labeled-only interval, the realized effective labeled sample size $n_B\,(SD_{\mathrm{labeled}}/SD_{\mathrm{PTD}})^2$, the closed-form effective sample size of ref.~\citep{Broskaetal2025} for the individually randomized Swedish trial, the PPI correlation $\tilde\rho$ with a Fisher-$z$ interval, the tuning weight and the prediction residual on split B.

For Ghana, where the labeled high-cash arm holds three vaccinated participants, the maximum-likelihood labeled fit separates in 37--40\% of replicates. We therefore report Firth-penalized labeled fits alongside (SI Appendix, section 2), and a wild cluster bootstrap with Webb weights as the registered robustness check \citep{CameronGelbachMiller2008, MacKinnonWebb2018}. % s6-main-methods-begin
The registered design-based coverage simulation for the Ghanaian intervals is reported in SI Appendix, section 7, at 500 datasets of 500 replicates each rather than the registered 2,000 of 2,000 (SI Appendix, section 2). Across combinations and arms, the prediction-powered and labeled-only intervals contain the true coefficient in 90--97\% and 91--97\% of the simulated datasets, and the wild cluster bootstrap interval, about half as wide on the observed data, in 77--88\%; the cluster-bootstrap intervals are therefore the ones we read.

\subsection*{What Training Teaches (Exploratory).}
These analyses follow a plan written before they were run and are not preregistered (SI Appendix, section 9). Each model is compared with a ridge-penalized logistic regression fitted to the same training data and reading the same participant profile. The pilot regression is fitted to split A on the arm and on the profile variables of the prompt, with the district in place of the community name in Ghana. The corpus regression is fitted to the 64,580 corpus records whose answer begins with Yes or No, on the country, the type of intervention, the type of outcome and the participant characteristics that the corpus shares with each target profile (15 in Sweden, 11 in Ghana). For the corpus-then-pilot models, a combined regression weights the two on split A. How much of a model's held-out predictions a regression reproduces is measured by regressing the former on the latter on the log-odds scale. Accuracy on the held-out sample is compared through differences paired over 2,000 bootstrap replicates (villages in Ghana) for treatment effects, by the root-mean-square error of the arm effects refitted in each replicate by fractional logit of the registered specification; for arm rates, by their mean absolute error; and for individual predictions, by the Brier skill score.

\subsection*{Exposure of the Training Data to the Target Trials.}
OLMo~2's training corpora are indexed in full. Before registration, we queried them for verbatim windows of each target paper, its identifiers and its data deposits through the infini-gram suffix-array API \citep{Liuetal2024}, across the pretraining mixture and the full training stacks of the 7B proxy and the 32B model. The Swedish paper is present (19 of 66 queries and 16 of 36 content windows return hits, all in pretraining); the Ghanaian paper, published after the corpora's effective cutoff, is absent (0 of 82). The fourth dimension is therefore registered as primary for the OLMo~2 models in Ghana and as illustrative elsewhere (SI Appendix, section 8). The probe detects verbatim exposure only, and GPT-4o and Qwen3 cannot be probed.

\subsection*{Robustness Checks and Subgroup Diagnostics.}
The plan registers five robustness checks: a fractional-logit fit on $\hat p_i$ in place of the draws; a pure-control reference in Ghana; evaluation sets enlarged by split B, and by splits A and B for the two training protocols that do not use the pilot; and split ratios of 20/30/50 and 20/10/70. It also registers subgroup diagnostics of the second and third dimensions by age band, gender, education and, in Ghana, region. All are reported in SI Appendix, section 6. Under every check the synthetic effects move by at most 1.5 pp in Sweden and 7 pp in Ghana, the released models keep the largest distances and the lowest skill, and the precision gains move with the sizes of the labeled and held-out samples, as the closed-form expression implies. In Ghana the enlarged evaluation sets and the pure-control reference move the human benchmark more than the synthetic estimates. The held-out high-cash villages vaccinate at 41\%, against 26\% in the trial as a whole, so the human high-cash effect is 11 pp on the held-out sample and about zero once split B is added, the same dependence on the sample that the sensitivity analysis above reveals.

\subsection*{Ethics and Preregistration.}
All sources in the training corpus and both target trials hold prior ethics approvals; this study uses only their public archives and generates no new human data. Models, training corpus, prompts, hyperparameters, splits, estimators and every reported indicator were fixed in a pre-analysis plan registered before data generation began \citep{duch_batistoni_beretta_low_2026}. The plan was registered on 28 June 2026 and amended twice through the registry's update mechanism (15 July and 8 September 2026), in each case before any confirmatory data generation; the analyses follow the amended plan, and SI Appendix, section 1 records the version history. Every departure from the plan is listed in SI Appendix, section 2 and treated as exploratory.

\subsection*{Data, Code and Models.}
Human data come from the original studies' public archives, and a datasheet documents every source and its redistribution status. The synthetic predictions, the code for training, inference and analysis,
every prompt and configuration file, and the LoRA adapters with their model
cards are released in the replication package, deposited on Zenodo
(doi:10.5281/zenodo.24530389); the adapters are additionally released on
Hugging Face.
}

\showmatmethods{} % Display the Materials and Methods section

\dataavail{The pre-analysis plan is registered on the OSF
(doi:10.17605/OSF.IO/68JFQ). The synthetic predictions, the training,
inference and analysis code, every prompt and configuration file, and the
datasheet are deposited on Zenodo (doi:10.5281/zenodo.24530389); the LoRA
adapters with their model cards are released on Hugging Face. Human trial
data are available from the original studies' archives \citep{Duchetal2023,
CamposMercadeetal2021}; the external-corpus sources and their access
conditions are listed in SI Appendix, section 3 and in the datasheet.}

\acknow{We are grateful for research funding from the University of Oxford ``Talking to Machines'' project and the Swiss National Science Foundation (grant 100018M-215519)}

\showacknow{} % Display the acknowledgments section

\bibsplit[3]

\bibliography{PAP,t2m_master}

\end{document}

%% file: Figures/Main/fig1_design.tex
% Figure 1: study design (TikZ). \input this file inside a figure* environment.
\begin{tikzpicture}[
  font=\sffamily\footnotesize,
  box/.style={draw, rounded corners=2pt, align=center, inner sep=4pt, minimum height=8mm, line width=0.5pt},
  cell/.style={draw, align=center, inner sep=3pt, minimum width=2.45cm, minimum height=1.25cm, line width=0.5pt},
  data/.style={box, fill=gray!12},
  model/.style={box, fill=blue!10},
  eval/.style={box, fill=orange!12},
  lab/.style={font=\sffamily\bfseries\small},
  arr/.style={-{Latex[length=2mm]}, line width=0.6pt},
  >=Latex
]
% ---------------- Panel A: training protocols ----------------
\node[lab, anchor=north west] (A) at (0,0) {A \ Training protocols};
\node[cell, fill=gray!8]   (s1) at (1.9,-2.05) {\textbf{T0}\\Released\\Model};
\node[cell, fill=green!10] (s2) at (4.5,-2.05) {\textbf{T1}\\External\\Corpus};
\node[cell, fill=cyan!12]  (s3) at (1.9,-3.45) {\textbf{T2}\\Pilot};
\node[cell, fill=violet!12] (s4) at (4.5,-3.45) {\textbf{T3}\\Corpus,\\Then Pilot};
\node[anchor=south, font=\sffamily\scriptsize, align=center] at (1.9,-1.3) {No External Corpus};
\node[anchor=south, font=\sffamily\scriptsize, align=center] at (4.5,-1.3) {External Corpus};
\node[anchor=east, font=\sffamily\scriptsize, align=right] at (0.6,-2.05) {No\\Pilot};
\node[anchor=east, font=\sffamily\scriptsize, align=right] at (0.6,-3.45) {Pilot};
% ---------------- Panel B: splits and uses ----------------
\node[lab, anchor=north west] (B) at (6.3,0) {B \ Splits of each trial and the two uses of the predictions};
\node[anchor=south, font=\sffamily\scriptsize, align=center, text width=2.0cm] at (7.4,-1.35) {Pilot from\\Target Trial};
\node[anchor=south, font=\sffamily\scriptsize, align=center, text width=2.0cm] at (9.5,-1.35) {Labelled Sample\\for PTD};
\node[anchor=south, font=\sffamily\scriptsize, align=center, text width=5.0cm] at (13.4,-1.35) {Held Out Data\\(treated as unlabelled sample in PTD)};
\draw[fill=cyan!12, line width=0.5pt]   (6.5,-1.95) rectangle (8.4,-1.45) node[midway, font=\sffamily\scriptsize] {A 20\%};
\draw[fill=yellow!35, line width=0.5pt] (8.4,-1.95) rectangle (10.3,-1.45) node[midway, font=\sffamily\scriptsize] {B 20\%};
\draw[fill=gray!10, line width=0.5pt]   (10.3,-1.95) rectangle (16.0,-1.45) node[midway, font=\sffamily\scriptsize] {C 60\%};
\node[eval, text width=4.3cm, anchor=north west] (fc) at (6.3,-2.55)
  {\textbf{Forecasting}\\Treatment effects (D1)\\Arm-level outcome rates (D2)\\Individual-level outcomes (D3)};
\node[eval, text width=4.3cm, anchor=north west] (es) at (11.2,-2.55)
  {\textbf{Augmenting estimation}\\Precision gain when predictions\\for split C join the observed\\outcomes of split B (D4)};
% ---------------- Panel C: one cell ----------------
\node[lab, anchor=north west] (C) at (0,-5.0) {C \ From training data to synthetic predictions};
\node[data, text width=2.6cm] (ext) at (1.55,-6.2) {External Corpus\\(8 experiments,\\6 surveys)};
\node[data, text width=2.6cm] (pil) at (1.55,-7.6) {Pilot\\(Split A of target trial)};
\node[model, text width=2.1cm] (ft) at (4.95,-6.9) {LoRA\\fine-tuning};
\node[model, text width=2.1cm] (mod) at (8.15,-6.9) {Synthetic\\participant};
\node[data, text width=3.0cm, anchor=north] (prompt) at (8.15,-8.0) {Prompt\\(Context -- Profile --\\Assigned Treatment)};
\node[eval, text width=4.0cm] (phat) at (12.2,-6.9) {$\hat p_i$: probability of vaccinating\\for each unit in the held out data};
\draw[arr] (ext) -- (ft);
\draw[arr] (pil) -- (ft);
\draw[arr] (ft) -- (mod);
\draw[arr] (prompt) -- (mod);
\draw[arr] (mod) -- (phat);
\end{tikzpicture}